\documentstyle[preprint,aps]{revtex}

\begin{document}

\draft

\title{Analysis of $\pi$d elastic scattering data to 500 MeV}
\author{Richard A. Arndt, Igor I. Strakovsky$^\dagger$ and Ron L. Workman}
\address{Department of Physics, Virginia Polytechnic Institute and State
University, Blacksburg, VA 24061}

\date{\today}
\maketitle

\begin{abstract}
An energy-dependent and a set of single-energy
partial-wave analyses of $\pi$d elastic scattering
data have been completed.  Amplitudes are presented for pion laboratory
kinetic energies up to 500~MeV.  These results are compared with
those found in other recent analyses.  We comment on the present database
and make suggestions for future work.

\end{abstract}
\vskip 0.3in

\narrowtext
\section{Introduction}
\label{sec:intro}

The $\pi d$ system is an important component of the more general $\pi NN$
problem.  At intermediate energies, the $\pi d$ and $N \Delta$ channels
provide much of the inelasticity associated with $pp$ scattering\cite
{ga90}.  We have previously analyzed the $pp\to pp$\cite{ar92}
and $\pi d\to pp$\cite{ar93} reactions. Here we give the results of our
analyses of $\pi d$ elastic scattering data.

A number of calculations have resulted in a
reasonable qualitative (and over some regions quantitative) description of
$\pi d$ scattering data\cite{ga90}.
The more theoretical approaches have had
difficulty in describing all observables\cite{ga90}, and have generally
concentrated on limited kinematic regions.
In addition, several partial-wave\cite{ar94}--\cite{hi94} analyses have
been performed. These analyses have
found some motivation from claims\cite{st91} of dibaryon resonances in $pp$
and $\pi d\to pp$ scattering reactions. Comparisons with data have been
complicated by the occasional appearance and subsequent disappearance of
sharp structures in some polarization observables.  We plan to examine these
questions through a coupled-channel analysis of $pp$ and $\pi d$ elastic
scattering combined with $\pi d\to pp$.
However, as a first step, we have analyzed $\pi d$ elastic scattering
separately.

The present analyses include measurements to 500~MeV in the laboratory
kinetic energy of the pion.  (This corresponds to laboratory kinetic energies
between 287.5~MeV and 1287~MeV in the $pp$ system.)  The value of $\sqrt{s}$
varies from 2.015~GeV to 2.437~GeV, and spans the range of energies typically
associated with dibaryon candidates.  This reaction displays the same
``resonancelike" behavior found in partial-wave analyses of $pp\to pp$ and
$\pi d\to pp$ scattering data.  The interpretation of this behavior generally
arouses strong reactions, either for or against the existence of intermediate
dibaryon resonances.  In the present work, our goal is simply a refined
understanding of the $\pi d$ elastic scattering amplitudes.  Interpretations
will depend upon the results of our larger joint analysis of the $pp$ and
$\pi d$ channels.

In the next section, we will make some comments on the database used in this
analysis. In Section~III, we will outline the general formalism for
$\pi$d elastic scattering.  Methods used in the partial--wave analysis will
be discussed in Section~IV.
Our main results will be presented in Section~V.  Here we will
also compare with the available data and other recent analyses.  Finally, in
Section~VI, we will summarize our findings and make some suggestions for
future investigations.

\section{The Database}
\label{sec:dat}

Experimental studies of $\pi$d elastic scattering began to produce
results in the 1950s.  At this time, the first measurements of total and
differential cross sections became available.  The trend of $\pi$d elastic
scattering data accumulation since 1952 is displayed in Fig.~1.  The rapid
increase in the number and type of measurements in the early 1980s was
motivated by a growing interest in the problem of exotics.  This reaction
was expected to give further information on the existence (or non-existence)
of dibaryon states suggested in analyses of $NN$ elastic scattering data.
Numerous high-quality deuteron polarization measurements were made.  The
total database more than doubled, and several partial-wave analyses
\cite{ar94}--\cite{hi94} were carried out at this time.  The present study
has utilized a set of data which is significantly larger,
and covers a broader energy interval.

Our total set of experimental data\cite{ak83}--\cite{we94} (1362 points)
includes measurements of the $\pi ^{+}$d (57) and $\pi ^{-}$d (67)
total cross sections ($\sigma _T$),  $\pi ^{+}$d (516) and
$\pi ^{-}$d (236) differential cross sections (d$\sigma$/d$\Omega$)
with unpolarized targets, the deuteron vector analyzing-power (iT$_{11}$)
for $\pi ^{+}$d (280) and $\pi ^{-}$d (5), the
deuteron tensor analyzing-power (T$_{20}$)
for $\pi ^{+}$d (42), the
combined deuteron tensor analyzing-powers ($\tau _{21}$ and $\tau _{22}$)
for  $\pi ^{+}$d  (123), and the laboratory
deuteron tensor analyzing-power (t$_{20}^{lab}$)
for $\pi ^{+}$d (30).  We have also included 6 unpolarized total
elastic cross sections ($\sigma _T ^{el}$).  Energy-angle
distributions for $d\sigma / d\Omega$, $iT_{11}$ and $T_{20}$
are given in Fig.~2.
Most of the $\pi ^{+}$d data are concentrated at low energies;
the $\pi ^{-}$d data tend to span a wider energy range.

As shown in Table I, we have removed $\sigma_T$ (2),
$d\sigma / d\Omega$ (88), $\sigma_{T}^{el}$ (3), iT$_{11}$
(6) and t$_{20}^{lab}$ (35) data corresponding to 9\% of the
total.  These measurements were the source of serious conflicts within the
database and were not included in the analysis.  For instance, many of the
t$_{20}^{lab}$ measurements (33 points) were produced by the Z\"urich
group\cite{ko83}. These were found to be in conflict with a number of
independent measurements
at LAMPF\cite{ho79,ho81,un85},
TRIUMF\cite{sm88,st89} and PSI\cite{ot88}.

\section{Formalism}
\label{sec:form}

The relations between partial-wave amplitudes and observables have been
given in a number of previous theoretical and phenomenological studies.
Many of the these relations are given in  the work of
Grein and Locher\cite{gr81}.  They have been included here for
completeness and to define our notation.
Due to parity conservation, there are 4 independent
helicity amplitudes for this reaction. Thus, for a reconstruction of the
scattering amplitude at fixed values of the energy and scattering angle,
one requires 7 independent measurements.  The amplitude,
$H_{\alpha \beta} (\theta )$, is labeled by the deuteron helicities
($\alpha$ and $\beta$) in initial and final states.  Here the angle
$\theta$ is the center-of-mass scattering angle of the outgoing pion.
Our notation for the helicity
amplitudes\cite{gr81} is given below:

\begin{eqnarray}
H_{++} \equiv H _1 & = &
\frac{1}{2} \sum \limits_{J \geq 1}{[(J + 1)~T ^J _{J-1~J-1} +
J~T ^J _{J+1~J+1} + (2~J + 1)~T ^J_{J~J} +
2~ \sqrt{J~(J + 1)} ~T ^J _{J-1~J+1}]~d _{11} ^J},
\nonumber \\
H_{+0} \equiv H _2 & = &
-\frac{1}{2} \sum \limits_{J \geq 1}{[\sqrt{2~(J + 1)}~(T ^J _{J+1~J+1} -
T ^J _{J-1~J-1}) + \sqrt{2}~T ^J _{J-1~J+1}]~d _{10} ^J},
\\
H_{+-} \equiv H _3 & = &
\frac{1}{2} \sum \limits_{J \geq 1}{[(J + 1)~T ^J _{J-1~J-1} +
J~T ^J _{J+1~J+1} - (2~J + 1)~T ^J _{J~J} +
2~ \sqrt{J~(J + 1)} ~T ^J _{J-1~J+1}]~d _{1-1} ^J},
\nonumber \\
H_{00} \equiv H _4 & = &
\sum \limits_{J \geq 0}{[J~T ^J _{J-1~J-1} +
(J + 1)~T ^J _{J+1~J+1} -
2 \sqrt{J~(J + 1)}~T ^J _{J-1~J+1}]~d _{00} ^J},
\nonumber
\end{eqnarray}
where the $d ^J _{\alpha \beta}$ are reduced rotation matrices.
The following symmetry relations

\begin{eqnarray}
H_{\alpha\; \beta} & = & (-1)^{\alpha + \beta} H_{-\alpha\; -\beta},
\\
H_{0 \; \beta} & = & - H_{\beta\; 0}
\nonumber
\end{eqnarray}
are also obeyed by the above helicity amplitudes.  The partial-wave
amplitudes, $T^J_{L^{\pi '}\;L^{\pi}}$, are labeled by the values of
$L^{\pi '}$ corresponding to the $\pi d$ final state, and $L^{\pi}$ for the
$\pi d$ initial state.  In the next section, and in our figures, we use the
notation $^3 L _J ^{\pi}$, and $\epsilon _J$ for the amplitudes
$^3(J-1)_J$ $-$ $^3(J+1)_J$ as in Ref.\cite{ar92}.

The various observables for $\pi$d elastic scattering are given
in terms of helicity amplitudes\cite{gr81,ob} below:

\begin{eqnarray}
I_{0} \equiv t_{00}^{00} & = & 2~\left| H _1 \right| ^2 +
4~\left| H _2 \right| ^2 + 2~\left| H _3 \right| ^2 +
\left| H _4 \right| ^2,
\\
\frac{d \sigma}{d \Omega} & = &  \sigma _g~I_{0},
\nonumber \\
\sigma _{T}^{el} & = & 4 \pi \sigma _g~\int \limits_{0}^{\pi}
{I_{0}\,\sin{\theta}\,d \theta},
\nonumber \\
\sigma _{T} & = & 4 \pi \sigma _g ~[2 ImH _1 (0) + ImH _4 (0)],
\nonumber \\
\sigma _{T}^{aligned} & = & 8 \pi \sigma _g ~[ ImH _1 (0) - ImH _4 (0)],
\nonumber \\
iT_{11} \equiv it_{11}^{00} & = & -\sqrt{6}~Im[H_2^ \ast (H _1 - H _3
+ H _4)] / I_{0},
\nonumber \\
T_{20} \equiv  t_{20}^{00} & = & \sqrt{2}~(\left| H _1 \right| ^2 -
\left| H _2 \right| ^2 + \left| H _3 \right| ^2 -
\left| H _4 \right| ^2) / I_{0},
\nonumber \\
T_{21} \equiv -t_{21}^{00} & = & -\sqrt{6}~Re[H_2^ \ast (H _1 - H _3
- H _4)] / I_{0},
\nonumber \\
T_{22} \equiv  t_{22}^{00} & = & \sqrt{3}~[2~Re(H_1^ \ast H _3) -
\left| H _2 \right| ^2] / I_{0},
\nonumber \\
t_{10}^{10} & = &  3~(\left| H _1 \right| ^2 - \left| H _3 \right| ^2) /
I_{0},
\nonumber \\
t_{11}^{10} & = & 3~Re[H_2^ \ast (H _1 + H _3)] / I_{0},
\nonumber \\
it_{21}^{10} & = &  3~Im[H_2^ \ast (H _1 + H _3)] / I_{0},
\nonumber \\
it_{22}^{10} & = &  3~\sqrt{2}~Im(H_1^ \ast H _3)] / I_{0},
\nonumber \\
t_{11}^{11} & = &  3~[Re(H_1^ \ast H _4) + \left| H _2 \right| ^2] /
I_{0},
\nonumber \\
t_{1-1}^{11} & = & 3~[-Re(H_3^ \ast H _4) + \left| H _2 \right| ^2] /
I_{0},
\nonumber \\
it_{20}^{11} & = &  -\sqrt{3}~Im[H_2^ \ast (H _1 - H _3 - 2~H _4)] /
I_{0},
\nonumber \\
it_{21}^{11} & = &  -3~Im(H_1^ \ast H _4) / I_{0},
\nonumber \\
it_{2-1}^{11} & = & -3~Im(H_3^ \ast H _4) / I_{0},
\nonumber \\
t_{22}^{11} & = &  3~\sqrt{2}~Im(H_2^ \ast H _1) / I_{0},
\nonumber \\
t_{2-2}^{11} & = & -3~\sqrt{2}~Im(H_2^ \ast H _3) / I_{0},
\nonumber \\
t_{20}^{20} & = &  (\left| H _1 \right| ^2 - 4~\left| H _2 \right| ^2 +
\left| H _3 \right| ^2 - 2~\left| H _4 \right| ^2) / I_{0},
\nonumber \\
t_{21}^{20} & = &  \sqrt{3}~Re[H_2^ \ast (H _1 - H _3 + 2~H _4)] /
I_{0},
\nonumber \\
t_{22}^{20} & = & \sqrt{6}~[Re(H_1^ \ast H _3) + \left| H _2 \right| ^2 ] /
I_{0},
\nonumber \\
t_{21}^{21} & = & 3~[Re(H_1^ \ast H _4) - \left| H _2 \right| ^2 ] /
I_{0},
\nonumber \\
t_{2-1}^{21} & = & 3~[Re(H_3^ \ast H _4) + \left| H _2 \right| ^2 ] /
I_{0},
\nonumber \\
t_{22}^{21} & = & 3~\sqrt{2}~Re(H_2^ \ast H _1) / I_{0},
\nonumber \\
t_{2-2}^{21} & = & -3~\sqrt{2}~Re(H_2^ \ast H _3) / I_{0},
\nonumber \\
t_{22}^{22} & = &  3~\left| H _1 \right| ^2 / I_{0},
\nonumber \\
t_{2-2}^{22} & = & 3~\left| H _3 \right| ^2 / I_{0}.
\nonumber
\end{eqnarray}

In these relations the factor $\sigma_g$ is equal to
${1\over 3}~ \bigl( { {\not h}\over k_{\pi} }\bigr)^2$, where $k_{\pi}$ is
pion momentum in the center-of-mass frame.  Here the $t$'s are spherical
harmonics\cite{bl85}.  The superscript *
denotes complex conjugation.  We have followed the Madison
convention\cite{gr81} in defining a coordinate system.
The observables
$d\sigma / d\Omega$, $\sigma _{T}^{el}$, and $\sigma_{T}$ are the usual
unpolarized cross sections.
$\sigma_T^{aligned}$ gives the spin-aligned total cross section.
The other observables involve either the polarization
of the initial or final deuteron ($iT_{11}$, $T_{20}$, $T_{21}$, and
$T_{22}$) or both.  The following relationships are valid due to
parity conservation and time reversal invariance:

\begin{eqnarray}
t_{L'M'}^{LM} & = & (-1)^{L + L' + M + M'}~t_{L'-M'}^{L-M},
\\
t_{L'M'}^{LM} & = & (-1)^{M + M'}~t_{LM}^{L'M'}.
\nonumber
\end{eqnarray}

We also require some combinations of deuteron tensor analyzing powers
measured by experimentalists. In Ref.\cite{ot881}, $\tau _{21}$ and
$\tau _{22}$ are defined by

\begin{eqnarray}
\tau _{22} & = & \sqrt{\frac{1}{6}}~T_{20} + T_{22},
\\
\tau _{21} & = & T_{21} + \frac{1}{2} ~ \tau _{22} =
{\frac{1}{2}} ~ \sqrt{\frac{1}{6}} ~T_{20} + T_{21} + \frac{1}{2} ~T_{22}.
\nonumber
\end{eqnarray}

The laboratory deuteron tensor analyzing-powers expressed\cite{ot881}
in terms of CM variables are

\begin{eqnarray}
t_{20}^{lab} & = & \frac{3 \cos^2 \theta _R - 1}{2}~T_{20} +
      2~\sqrt{\frac{3}{2}} \sin{\theta _R} \cos{\theta _R}~T_{21} +
        {\frac{3}{2}} \sin^2 \theta _R~T_{22},
\\
t_{21}^{lab} & = & \sqrt{\frac{3}{2}} \sin{\theta _R} \cos{\theta _R}~
       T_{20} + (1 - 2~\cos^2 \theta _R)~T_{21} -
        \sin{\theta _R} \cos{\theta _R}~T_{22},
\nonumber \\
t_{21}^{lab} & = & {\frac{1}{2}} \sqrt{\frac{3}{2}} \sin^2 \theta _R~
       T_{20} - \sin{\theta _R} \cos{\theta _R}~T_{21} +
        {\frac{1 + \cos^2 \theta _R}{2}}~T_{22},
\nonumber
\end{eqnarray}
where $\theta _R$ is the deuteron recoil angle in the lab frame.

\section{Partial-Wave Analysis}
\label{sec:pwa}

The energy-dependence of our global fit was obtained through a
coupled-channel K-matrix form, in order to ensure that unitarity
would not be violated. The ``inelastic'' channel was non-specific
but included to account for the coupled $pp$ and $N\Delta$ reactions.
For single-channel $\pi d$ states (for example, $^3 D_2$) this resulted
in a 2x2 matrix, and for the spin-coupled states (for example,
$^3P_2$, $\epsilon_2$, $^3F_2$) we had a 3x3 matrix. The
matrix elements were expanded as polynomials in the pion laboratory
energy, and an appropriate phase-space factor was included in the
$\pi d$ elastic elements to ensure proper threshold behavior.
This analysis included 21 searched partial-waves and
66 varied parameters.  Amplitudes with $J\le 5$ were considered.
The solution gave a $\chi^2$ of 2743 for the 1362 data
and 333 experiments below 500~MeV.
Coulomb modifications of the phase-space factors were attempted but
discarded in the final fit; the sensitivity to such refinements appeared
minimal.

Single-energy solutions were
produced up to 300~MeV, using a binning width of
10$\pm$5~MeV.  We used mainly the energy values chosen in previous
single-energy PWAs\cite{ar94}--\cite{ga89}.  Starting values for the
partial-wave amplitudes, as well as their (fixed) energy derivatives, were
obtained from the energy-dependent fit.  The scattering database was
supplemented with a constraint on each varied amplitude.  Constraint
errors were taken to be 0.02.  This was added, in quadrature, to 5\% of
the amplitude.  Such constraints were essential to prevent the solutions
from ``running away'' when the bin was relatively empty of scattering data.
These errors were generous
enough that they afforded little constraint for those solutions where
sufficient data existed within the bin.

Single-energy analyses are done in order to reveal ``structure'' which may
be missing from the energy-dependent fit.
Little compelling evidence was found for such structure.
Results of the single-energy analyses are
summarized in Table~II.  A maximum of
9 partial-waves ($^3$S$_1$, $\epsilon_1$, $^3$P$_{0, 1, 2}$,
$^3$D$_{2, 3}$, and $^3$F$_{3, 4}$) were searched in the single-energy
analyses. The remaining partial-waves were fixed at the energy-dependent
values.

Several Coulomb correction schemes were tested. Our results
were found to be relatively independent of the chosen form. Differences
are most apparent in charge-asymmetry observables, as shown in
Ref.\cite{frohl}. In the present analysis, we have adopted Hiroshige's
formulation\cite{hi94}.

\section{Results and Comparisons}
\label{sec:rs}

Results for the partial-wave amplitudes, defined in Eq.~(1), are
shown in Fig.~3. Over our energy range, the dominant amplitudes are the
$^3$S$_1$, $^3$P$_1$, $^3$P$_2$, $^3$D$_2$, $^3$D$_3$, and
$^3$F$_4$. The compatibility of our energy-dependent and single-energy
solutions is also evident in Fig.~3. The lack of single-energy solutions
beyond 300 MeV is due to the data distribution displayed in Fig.~2.

As mentioned in Section~IV, the energy-dependent solution gives a reasonable
overall fit to the data.  While the $\chi^2$/datum was about 2 for the
selected data base, a much higher value would have resulted from the
total set of measurements.  Some of the
data conflicts are apparent in Fig.~4, where we have given predictions for
observables at $T_{\pi ^{+}}$ = 256~MeV. The data are generally well
described at this energy, except possibly the most forward $iT_{11}$
measurements. We should mention that $\pi^- d$ measurements account
for only about a quarter of the total dataset. In addition, there is a
noticable difference in the $\chi^2$/datum (1.86 for $\pi^+ d$ versus
2.55 for $\pi^- d$). Various $\pi^+ d$ total
cross sections are plotted in Fig.~5.

We have also compared our results to those from several other groups.
The last single-energy partial-wave analyses for $\pi$d elastic scattering
were published by the Grenoble--Rehovot\cite{ar94}, Osaka\cite{hi90},
Saskatoon\cite{st87}, and Mexico--Karlsruhe\cite{ga89} groups.  These
results have generally covered a more narrow energy interval
(7 points from 82 to 292~MeV in\cite{ar94}, 8 points from
114 to 325~MeV in\cite{hi90}, 9 points from 117 to 324~MeV in\cite{st87},
and 2 points at 256 and 294~MeV in \cite{ga89}).  The energy-dependent
analysis of Ref.\cite{hi94} covered the region between 65~MeV and
275~MeV. We have plotted our energy-dependent results along with several
analyses in Fig.~6. The figure displays the experimental
version of model (iv) from Ref.\cite{ga89}, and the version of Ref.\cite{st87}
using Blankleider-Afnan\cite{bl85} model amplitudes as input.
The agreement is qualitative at best.

\section{Summary and Conclusions}
\label{sec:sc}

In this work we have analyzed a $\pi$d elastic scattering data base which is
significantly larger than those used in previously published analyses.  A
reasonable fit to this database was found.  While there are considerable
differences between the results of this and previous analyses, one common
qualitative feature is present. Several dominant partial-wave amplitudes
display a ``resonancelike'' behavior. The correlation between these
amplitudes is particularly evident in the Argand
plot of Fig.~7. This behavior is very similar to that found in our
recent analyses of $\pi d\to pp$\cite{ar93} and $pp$\cite{ar92} elastic
scattering data.

The present work is completely free of model-based constraints.
Previous analyses have generally used theory as a guide where
insufficient experimental information was available. This factor is
a likely source for some of the discrepancies evident in Fig.~6.
The evolving data base provides another.

While we have analyzed data to 500 MeV, the results above 300 MeV should
not be taken too seriously. Most of the data base is concentrated below
300 MeV, with only about 200 measurement covering the region between
300 MeV and 500 MeV. Clearly, we require much
additional (and more consistent) data to define a unique solution
(both above and below 300 MeV).
We should mention that the present solution predicts a rich
structure for many of the spin-transfer observables defined in
Eqs.~(3). While only a few of these quantities have been
measured, some new PSI
measurements\cite{br94} of
t$_{11}^{11}$,  it$_{20}^{11}$, it$_{21}^{11}$, and  t$_{22}^{11}$ at
134, 180, and 219~MeV will soon be available.

This reaction is now incorporated into the SAID program\cite{tel}, which is
maintained at Virginia Tech.  Detailed information regarding the database,
partial-wave amplitudes and observables may be obtained either interactively,
through the SAID system (for those who have access to TELNET), or directly
from the authors.

\acknowledgments

The authors express their gratitude to J.L. Beveridge, B. Brinkm\"oller,
D.V. Bugg, J.-P. Egger, R.R. Johnson, R. Meier, R.C. Minehart, R.J. Peterson,
R.A. Ristinen, B. Saghai, G.R. Smith, N.R. Stevenson, Y. Sumi, and R. Tacik
for providing experimental data prior to publication and clarification of
information already published.  We are grateful to D.V. Bugg, N. Hiroshige,
G. Jones, M.P. Locher, D.F. Measday, T. Mizutani, A. Rinat, R.A. Ristinen,
and G.R. Smith for providing available partial--wave solutions and for
helpful information concerning the various Coulomb correction schemes.
I.S. acknowledges the
hospitality extended by the Physics Department of Virginia Tech.  This work
was supported in the part by the U.S.~Department of Energy Grant
DE--FG05--88ER40454.


\newpage
{\Large\bf Figure captions}\\
\newcounter{fig}
\begin{list}{Figure \arabic{fig}.}
{\usecounter{fig}\setlength{\rightmargin}{\leftmargin}}
\item
{Data accumulation from 1952 to the present.  Arrows indicate the year
when measurements of particular observables were first published.}
\item
{Energy-angle distribution of total dataset.
(a)~differential cross section, $d\sigma / d\Omega$ for $\pi ^{+}$d,
(b)~differential cross section, $d\sigma / d\Omega$ for $\pi ^{-}$d,
(c)~deuteron vector analyzing-power, $iT_{11}$ for $\pi ^{+}$d,
(c)~deuteron tensor analyzing-power, $T_{20}$ for $\pi ^{+}$d.}
\item
{Partial-wave amplitudes from 0 to 500 MeV. Solid curves are the real
parts of amplitudes; dashed curves are the imaginary parts. Single-energy
solutions are plotted as black circles (real part) and open circles
(imaginary part). All amplitudes have been multiplied by a factor
of 10$^3$, and are dimensionless.}
\item
{Predictions for $\pi^{+}$d observables at $T_{\pi}$ = 256~MeV.  Data
have been normalized.
(a)~$d\sigma / d\Omega$,    (b)~$iT_{11}$,    (c)~$T_{20}$,
(d)~$T_{21}$, (e)~$T_{22}$, (f)~$\tau _{21}$, (g)~$\tau _{22}$,
(h)~$t_{20}^{lab}$.}
\item
{Total cross sections, $\sigma$, for $\pi^+ d$ scattering. (a)~unpolarized
total cross section $\sigma_T$ (solid curve).
(b)~contribution\cite{ar93} from $\pi^+ d\to pp$ $\sigma^{pp}_T$ (dashed
curve),
(c)~the total elastic cross section $\sigma^{el}_T$ (dotted curve), and
the remainder $\Delta \sigma$ given by
$\sigma_T - \sigma^{el}_T - \sigma^{pp}_T$.}
\item
{Plots of the amplitudes (a)~$^3$S$_1$, (b)~$^3$P$_1$,
(c)~$^3$P$_2$, (d)~$^3$D$_2$, (e)~$^3$D$_3$, and (d)~$^3$F$_4$.
The solid (dashed) curves
give the real (imaginary) part from the present energy-dependent solution.
The circles, squares, triangles, and diamonds are results from
Refs.\cite{hi94}, \cite{st87}, \cite{ga89}, and \cite{ar94}, respectively.
Filled symbols give real parts; open symbols give imaginary parts.  All
amplitudes have been multiplied by a factor of $10^3$, and are
dimensionless.}
\item
{Argand plot of the  dominant $\pi$d partial-wave amplitudes $^3$P$_2$,
$^3$D$_3$, and $^3$F$_4$ which correspond to the $^1$D$_2$, $^3$F$_3$, and
$^1$G$_4$ pp states, respectively.  (Compare Fig.~7 of reference\cite{ar93}).
The ``X" points denote 50~MeV steps.  All amplitudes
have been multiplied by a factor of $10^3$.}
\end{list}

\newpage
\mediumtext

TABLE I. Number and type of observables used in the present analysis to
500~MeV in the pion laboratory kinetic energy.  The number of excluded
data is also given.
$^a$See Refs.\cite{br63} and \cite{ig55}.
$^b$See Refs.\cite{ga73}, \cite{mi81}, \cite{ko93}, and \cite{ma84}.
$^c$See Refs.\cite{ak83}, \cite{br63}, \cite{ma84}, and \cite{na55}.
$^d$See Refs.\cite{br63}, \cite{na55}, and \cite{pe63}.
$^e$See Ref.\cite{sm84}.
$^f$See Ref.\cite{ko94}.
$^g$See Refs.\cite{ko83} and \cite{st89}.
\vskip 10pt
\centerline{
\vbox{\offinterlineskip
\hrule
\hrule
\halign{\hfill#\hfill&\qquad\hfill#\hfill&\qquad\hfill#\hfill
&\qquad\hfill#\hfill&\qquad\hfill#\hfill\cr
\noalign{\vskip 6pt}
&$\pi^{+}$d&&$\pi^{-}$d&\cr
\noalign{\vskip 6pt}
\noalign{\vskip 6pt}
Observable&No. of Data&Deleted data&No. of Data&Deleted data\cr
\noalign{\vskip 6pt}
\noalign{\vskip 6pt}
\noalign{\hrule}
\noalign{\vskip 10pt}
$\sigma _{T}$&57&0&69&$2^a$\cr
\noalign{\vskip 6pt}
d$\sigma$/d$\Omega$&572&$56^b$&268&$32^c$\cr
\noalign{\vskip 6pt}
$\sigma _{T}^{el}$&3&0&6&$3^d$\cr
\noalign{\vskip 6pt}
$iT _{11}$&285&$5^e$&6&$1^f$\cr
\noalign{\vskip 6pt}
$T _{20}$&42&0&0&0\cr
\noalign{\vskip 6pt}
$\tau _{21}$&47&0&0&0\cr
\noalign{\vskip 6pt}
$\tau _{22}$&76&0&0&0\cr
\noalign{\vskip 6pt}
$t _{20}^{lab}$&65&$35^g$&0&0\cr
\noalign{\vskip 6pt}
\noalign{\hrule}
\noalign{\vskip 6pt}
Total&1147&96&349&38\cr
\noalign{\vskip 10pt}
No. of Energies&239&16&121&11\cr
\noalign{\vskip 10pt}}
\hrule}}
\vfill
\eject
Table II. Single-energy (binned) fits and $\chi^2$ values.
$N_{prm}$ is the number of amplitudes (real + imaginary) varied in the fit.
$\chi^2_C$ is due to the amplitude constraints, $\chi^2_D$ is the
contribution from data,  and $\chi^2_E$ is given by the energy-dependent
fit, SM94.
\vskip 10pt
\centerline{
\vbox{\offinterlineskip
\hrule
\hrule
\halign{\hfill#\hfill&\qquad\hfill#\hfill&\qquad\hfill#\hfill
&\qquad\hfill#\hfill&\qquad\hfill#\hfill&\qquad\hfill#\hfill
&\qquad\hfill#\hfill\cr
\noalign{\vskip 6pt} %
$T_{\pi}$&Range&$N_{data}$&$N_{prm}$&$\chi^2_C$&$\chi^2_D$&$\chi^2_E$\cr
 (MeV) &      (MeV)      &     &    &      &        &       \cr
\noalign{\vskip 6pt}
\noalign{\hrule}
\noalign{\vskip 10pt}
  65 &  $58.0- 72.0$ &  54 &  4 &  2.4 &  86.6 &  129.9\cr
\noalign{\vskip 6pt}
  87 &  $72.0- 85.5$ &  24 &  8 &  0.4 &   20.4 &   30.3\cr
\noalign{\vskip 6pt}
 111 & $107.5-125.2$ &  82 &  8 &  0.7 &   68.9 &   68.9\cr
\noalign{\vskip 6pt}
 125 & $115.0-134.0$ & 170 &  12 &  5.1 &  154.1 &  205.0\cr
\noalign{\vskip 6pt}
 134 & $124.0-142.8$ & 258 & 12 &  4.2 &  293.6 &  362.1\cr
\noalign{\vskip 6pt}
 142 & $133.0-152.0$ & 284 & 14 &  8.1 &  345.4 &  442.6\cr
\noalign{\vskip 6pt}
 151 & $141.0-160.6$ & 154 & 16 &  11.0 &  186.9 &  256.3\cr
\noalign{\vskip 6pt}
 182 & $174.0-189.5$ & 168 & 16 &  19.6 &  300.3 &  445.8\cr
\noalign{\vskip 6pt}
 216 & $206.0-220.0$ &  99 & 18 &  1.6 &  121.3 &  148.1\cr
\noalign{\vskip 6pt}
 230 & $220.0-238.0$ &  53 & 18 &  7.8 &   51.3 &   96.5\cr
\noalign{\vskip 6pt}
 256 & $254.0-260.0$ & 125 & 18 &  2.9 &  132.5 &  200.6\cr
\noalign{\vskip 6pt}
 275 & $270.5-284.4$ &  40 & 18 &  2.9 &   17.4 &   42.6\cr
\noalign{\vskip 6pt}
 294 & $284.4-300.0$ & 132 & 16 &  3.3 &  228.5 &  263.6\cr
\noalign{\vskip 10pt}}
\hrule}}

\begin{references}
\bibitem{tel}
Those with access to TELNET can run the SAID program with a link to
\hfil\break
VTINTE.PHYS.VT.EDU.  The login/password is: PHYSICS/QUANTUM.  The user may
view the current database and compare our solutions to those of other groups.
\end{references}
\end{document}